\newcommand{\pmdg}[2]{\makebox[1cm][r]{#1}$\pm$\makebox[1cm][r]{#2}}

\newcommand{\beq}{\begin{equation}}
\newcommand{\eeq}{\end{equation}}
\newcommand{\beqn}{\begin{eqnarray}}
\newcommand{\eeqn}{\end{eqnarray}}

\newcommand{\dps}{\mbox{dPS}}
\newcommand{\smc}{\scriptsize}

\documentstyle[11pt,citesort,ifthen]{article}

\begin{document}

\thispagestyle{empty}

\newcommand{\dgemail}{{\it Electronic
mail address: {\tt Dirk.Graudenz\char64{}psi.ch}}}

\newcommand{\dgwww}{{\it WWW URL:}
{\tt http://www.hep.psi.ch/graudenz/index.html}}

\renewcommand{\thefootnote}{\fnsymbol{footnote}}
\setcounter{footnote}{0}

\hspace*{\fill}
\begin{minipage}[t]{3cm}
\footnotesize
November 1997
\end{minipage}
\vspace{1.0cm}

\begin{center}

{\LARGE\bf {\tt DISASTER++}
}\\ 
\vspace{5mm} 
{\Large Version 1.0}

\vspace{1.0cm}

{\bf Dirk Graudenz}$\;$\footnote[1]{\dgemail}$\;$\footnote[5]{
\dgwww}\\
\vspace{0.1cm} 
{\it Paul Scherrer Institut\\
5232 Villigen PSI, Switzerland\\   
}

\vspace{0.5cm}   

\begin{minipage}{14cm}   

\begin{center} {\bf Abstract}\end{center}

{
\small
{\tt DISASTER++} is a {\tt C++} class library for the calculation of
(1+1) and (2+1)-jet-like quantities in deeply inelastic lepton--nucleon
scattering for one-photon exchange in next-to-leading-order QCD perturbation 
theory. The calculation is based on the subtraction formalism. The user
has access to an event record such that an arbitrary set of infrared-safe 
observables can be calculated in a single run. Compared to other existing 
universal programs, the full dependence on the number of flavours and on the 
renormalization and factorization scales is made explicit. An interface class
providing a simple interface from {\tt C++} to existing {\tt FORTRAN}
programs is available. In a preliminary study {\tt DISASTER++} is compared 
to two other programs for various bins of the lepton variables $x_B$ and $y$, 
where a particular emphasis is put on different behaviours for
$\xi \rightarrow 1$ of the parton densities $f(\xi)$. We find good agreement
of {\tt DISASTER++} and {\tt DISENT} (Version 0.1). The comparison of 
{\tt DISASTER++} and {\tt MEPJET} (Version 2.0) leads to several discrepancies.
}

\end{minipage}

\vspace{0.5cm}

\end{center}

\renewcommand{\thefootnote}{\arabic{footnote}}
\setcounter{footnote}{0}

\clearpage

\section{Introduction}
For studies of the hadronic final state in high-energy collisions, 
versatile programs for the calculation of QCD corrections are required.
The extraction of scale-dependent 
physical quantities such as the running strong coupling
constant $\alpha_s\left(Q^2\right)$ and parton densities
$f_i\left(\xi,Q^2\right)$ requires precise predictions 
in next-to-leading order of QCD perturbation theory.
At the electron--proton collider HERA at DESY in Hamburg, 
the strong coupling constant has been measured via jet rates 
\cite{1,2}. There is also a direct fit of the gluon density 
$f_g(\xi, Q^2)$ \cite{3} based on a Mellin transform
method \cite{4,5}. Calculations for jet cross sections 
in deeply inelastic scattering for the case of the modified JADE 
scheme
have been performed \cite{6,7,8,9,10}\footnote{
In these calculations, terms of the form $c\log c$, 
$c$ being the jet cut, have been neglected. This implies in particular
a certain insensitivity with respect to the jet recombination scheme.
The set-up of the calculations \cite{6,7,9} is such that
a jet consisting of two partons is always mapped onto a massless jet. 
Therefore the jet definition scheme which is used on experimental data
should be a ``massless'' scheme (this excludes, for example, the E-scheme).
The variation of jet cross sections within the possible massless schemes
cannot be modelled by that calculation.}
and implemented in the 
two programs {\tt PROJET} \cite{11} and {\tt DISJET} \cite{12}.

In the meantime,
calculations for arbitrary infrared-safe observables in 
deeply inlastic scattering have become available \cite{13,14}.
In the last few years, the technology for the calculation
of QCD corrections in next-to-leading order
has developed considerably. 
The main problem in higher-order QCD calculations is the occurence of
severe
infrared singularities (they ultimately cancel for infrared-safe
observables, or are absorbed into process-independent, physical
distribution functions such as parton densities and fragmentation functions).
There are explicit algorithms available
which permit the calculation to be done in a ``universal'' way: the 
infrared singularities are subtracted such that arbitrary 
infrared-safe observables can be calculated numerically. In principle, 
all existing algorithms are variations on a common theme, namely the
interplay of the factorization theorems of perturbative QCD and the 
infrared-safety of the observables under consideration.
There are two different ways to achieve the separation of
divergent and finite contributions:
the phase-space-slicing method \cite{15} and
the subtraction method \cite{16}. 
Both methods have their merits and drawbacks.
\begin{description}
\item[{\rm\unboldmath ($\alpha$)}] 
The phase-space-slicing method relies on a separation of
singular phase-space regions from non-singular ones by means of a 
small slicing parameter $s\rightarrow 0$. The divergent parts are evaluated 
under the assumption that terms of ${\cal O}(s (\log s)^n)$ can be dropped.
The analytically evaluated phase-space integrals yield terms of the form
$(\log s)^m$, which explicitly cancel against equivalent terms of opposite
sign from a numerically performed phase-space integration.
The simplicity of this scheme is obvious.
The main problem is the residual dependence on the technical cut 
parameter~$s$
(in practice the limit $s\rightarrow 0$ is not checked for every observable, 
but it is assumed that a fixed small value will be sufficient).
Moreover, the numerical cancellation of the logarithmic terms by means
of a Monte-Carlo integration is delicate.
There is a calculational scheme available for the determination of 
the explicit phase space integrals over the singular regions \cite{17}. 
For initial and final-state hadrons this scheme moreover 
requires the introduction
of so-called {\it crossing functions} \cite{18}, 
to be evaluated for every parton density parametrization.
For deeply-inelastic lepton--nucleon scattering, an implementation 
of this calculational scheme is provided by Mirkes and Zeppenfeld
in {\tt MEPJET} \cite{19}.
\item[{\rm\unboldmath($\beta$)}] 
The subtraction method is technically more involved, 
since the infrared singularities are cancelled point-by-point in 
phase space. The subtraction terms have, owing to the factorization
theorems of perturbative QCD, a simple form. The problem is 
to arrange the subtractions in such a way that in the numerical evaluation
no spurious singularities appear. A general framework, using a specific
phase space mapping besides the factorization theorems, is given by Catani
and Seymour in  Ref.~\cite{20}, and implemented in {\tt DISENT} 
\cite{21}.

The approach of the present 
paper is to use a generalized partial fractions
formula to separate the singularities \cite{22}. The method is
briefly explained in Section~\ref{algorithm}. We will describe in some
detail the implementation {\tt DISASTER++}\footnote{
This is an acronym for ``Deeply Inelastic Scattering: All Subtraction Through
Evaluated Residues''.}
in the form of a {\tt C++} class library.

There are two reasons for a new calculation.
(a) The existing
programs have the restriction that the number of flavours is fixed  
($N_f=5$ in the case of {\tt MEPJET}
and $N_f$ fixed, but arbitrary for {\tt DISENT}).
For studies of the scale-dependence it is
necessary to
have a variable number of flavours,
in order to be consistent with the scale evolution
of the strong coupling constant and the parton densities.
{\tt DISASTER++} makes the $N_f$ dependence explicit in the ``user routine''
on an event-by-event basis,
and thus results for several renormalization and factorization scales
can be calculated simultaneously.
(b) {\tt DISASTER++}
is already set up such that the extension to one-particle-inclusive   
processes will be possible without the necessity of re-coding
the contributions which are already present for
the jet-type observables. This option will be made available
in future versions of the program, as soon as the remaining contributions
for one-particle-inclusive processes are implemented.

\end{description}

The outline of this paper is as follows. In Section~\ref{algorithm}
we briefly review the algorithm employed in the present calculation.
In Section~\ref{structure} the {\tt FORTRAN} interface
to the {\tt C++} class library is described. 
Some remarks concerning the installation of the package are made
in Section~\ref{installation}.
A comparison of the available universal programs 
{\tt DISASTER++} (Version 1.0), {\tt MEPJET} (Version 2.0) 
and {\tt DISENT} (Version 0.1)
is presented in 
Section~\ref{comparison}.
In a previous version of this paper, we have drawn 
the conclusion that we find an overall, but not completely satisfactory
agreement of {\tt DISASTER++} and {\tt MEPJET}, and 
that there are large deviations when comparing
{\tt DISASTER++} and {\tt DISENT}.
One of the purposes of this paper is to present the results of a comparison
of {\tt DISASTER++} and a new, corrected version (0.1) 
of {\tt DISENT}. We now find
good agreement of the two programs.
We also give a few more results for {\tt MEPJET}, in particular
for the dependence on the technical cut~$s$. It turns out that
even for very small values of~$s$ 
we do not achieve agreement with the
{\tt DISASTER++}~/ {\tt DISENT} results
for several cases under
consideration\footnote{
In a very recent paper \cite{23}, E.~Mirkes quotes the results of the
comparison of the three programs as performed in the 
previous version of this paper \cite{24} as resulting in 
a ``so far satisfactory agreement''. This is a 
misquotation. The formulation in Ref.~\cite{24} was that 
for {\tt MEPJET} and {\tt DISASTER++} we find an ``overall, though not 
completely satisfactory agreement'', and that the results of {\tt DISENT}
(Version 0.0) ``differ considerably''. Moreover, in the summary
of Ref.~\cite{24} we mention that a few deviations of {\tt MEPJET} and
{\tt DISASTER++} are present. We wish to stress that there is a certain 
semantic
gap between the expression 
``satisfactory agreement'' and the results just quoted.
}.
The paper closes with a summary.
The contents of this paper are mainly technical. The details of the calculation
and phenomenological applications will be described in a forthcoming 
publication.

\section{The Algorithm}
\label{algorithm}
The calculation is based on the subtraction method. A simple example
to illustrate this method in general, and a comparison 
with the phase-space-slicing
method, is given in Ref.~\cite{25}.
For a more detailed exposition of the contents of this section, 
see Ref.~\cite{22}.

The subtraction method is one of the solutions for the problem of 
how to 
calculate 
numerically 
infrared-safe observables without having 
to modify the calculation for every observable under consideration.
In QCD calculations, infrared singularities cancel for sufficiently 
inclusive observables. 
The factorization theorems of perturbative
QCD (see Ref.~\cite{26} and references therein) 
together with the infrared-safety of the observable under consideration
guarantee that 
the structure of the limit of the convolution of the parton-level cross 
section with the observable in soft and collinear regions of phase space
is well-defined and factorizes in the form of a product of a kernel 
and the Born term.
This property allows, for the real corrections, the definition of a subtraction 
term for every phase-space point.
Formally:
\begin{eqnarray}
\int\dps^{(n)}\,\sigma\,{\cal O}
&=& \sum_A \int \dps_{i_A} \,k_A \left(
\int \dps^{(n-1)} \tau_A -
 \left[
   \int \dps^{(n-1)} \tau_A
 \right]_{\mbox{\smc soft/coll.~limit}}
\right)\nonumber\\
&+& \sum_A \int \dps_{i_A} \,k_A \left[
   \int \dps^{(n-1)} \tau_A
    \right]_{\mbox{\smc soft/coll.~limit}},
\end{eqnarray}
where $\sigma$ is the parton-level cross section, $\cal O$ is the
infrared-safe observable, $k_A$ is a singular kernel, 
and $\tau_A$ is the non-singular part of the product $\sigma\,\cal O$.
The index~$A$ runs over all possible soft, collinear and 
simultaneously soft and collinear singularities of~$\sigma$.
The first integral is finite and can be calculated numerically. The second
integral contains all infrared singularities. The term in the square bracket
has a simple structure
because of the factorization theorems of QCD, and the one-particle
integral over the kernel $k_A$ and the factorization contribution from the
term in the square brackets can be performed easily.
This subtraction formula works only if the subtraction terms do not
introduce spurious singularities for the individual terms that eventually
cancel in the sum. This is achieved by a separation of all singularities
by means of a general partial fractions formula
\beq
\label{pfid}
\frac{1}{x_1\,x_2\cdots x_n}
=\sum_{\sigma\in S_n}
\frac{1}{x_{\sigma_1}\,(x_{\sigma_1}+x_{\sigma_2})\cdots
         (x_{\sigma_1}+\ldots+x_{\sigma_n})},
\eeq
where the sum runs over all $n!$ permutations of $n$~objects.

In {\tt DISASTER++}, the processes for (1+1) and (2+1)-jet production 
for one-photon exchange are implemented. The program itself, however, 
is set up in a much more general way. The implemented subtraction procedure 
can handle arbitrary number of final-state partons, and zero or one incoming 
partons (an extension to two incoming partons is possible). The {\tt C++}
class library is intended to provide a very general framework for 
next-to-leading-order QCD calculations for arbitrary 
infrared-safe observables. Of course, the explicit matrix 
elements (Born terms, virtual corrections and factorized real corrections)
have to be provided for every additional process to be included.

\section{Program Structure}
\label{structure}
We now describe the {\tt FORTRAN} interface to the {\tt C++}
class library. The {\tt C++} user interface will be documented in a 
forthcoming extension of this manual.

To set the stage, let us first introduce some terminology.
The user has to provide several subroutines which are called by 
{\tt DISASTER++} for every generated event. Each {\bf event} 
$e_n$, $n=1\ldots N$
consists of a set of
{\bf phase spaces} ${\cal P}_{nr}$, $r=1\ldots R_n$, 
and a set of {\bf contributions} ${\cal C}_{ni}$, 
$i=1\ldots L_n$. Phase spaces $\cal P$
provide a set of four-vectors of initial and final-state
particles, which are used to calculate observables
${\cal O}({\cal P})$.
Contributions ${\cal C}_{ni}$ consist of a list of 
{\bf weights} $w_{nij}$, $j=1\ldots K_{ni}$ (here: 
$K_{ni}=11$) which have to be multiplied
by certain {\bf flavour factors} $F_{nij}$. 
Every contribution ${\cal C}_{ni}$ has an associated
phase space ${\cal P}_{nr_{ni}}$; 
it is generally the case that particular phase spaces are 
used for different contributions. Flavour factors are products
of parton densities, quark charges, powers of the strong coupling constant, 
and powers of the electromagnetic coupling constant.

The expectation value $\langle {\cal O} \rangle$ 
of a particular observable is given by the following 
sum:
\begin{equation}
\label{exval}
\langle {\cal O} \rangle = 
   \sum_{n=1}^N
   \sum_{i=1}^{L_n}
   {\cal O}({\cal P}_{nr_{ni}})    
   \sum_{j=1}^{K_{ni}}
   w_{nij} F_{nij}.
\end{equation}
The first sum is the main loop of the Monte Carlo integration.

\noindent
The user has to provide a subroutine
{\tt user1} and
a function 
{\tt user2}.
The subroutine
{\tt user1(iaction)} is called from {\tt DISASTER++} in the following cases:
\begin{description}
\item{\quad{\tt iaction=1}:} {\ }\\after start-up of {\tt DISASTER++}
\item{\quad{\tt iaction=2}:} {\ }\\before the end of {\tt DISASTER++}
\item{\quad{\tt iaction=3}:} {\ }\\before the start of the grid-definition 
run of the adaptive Monte-Carlo routine, or before the final run of
the adaptive integration, in case that there is no grid-definition run
\item{\quad{\tt iaction=4}:} {\ }\\before the start of the final
run of the adaptive Monte-Carlo routine
\item{\quad{\tt iaction=5}:} {\ }\\after the final
run of the adaptive Monte-Carlo routine
\item{\quad{\tt iaction=6}:} {\ }\\once for every event (to initialize intermediate 
weight sums, etc.)
\item{\quad{\tt iaction=7}:} {\ }\\signals that the event has to be dropped
for technical reasons
\end{description}

\noindent
The function {\tt user2(...)} is called from {\tt DISASTER++}
after an event has been constructed.
It has the following arguments (in an obvious
notation):
\begin{verbatim}
      double precision function
     &                 user2(
     &                   integer nr, 
     &                   integer nl,
     &                   double precision fvect(0..3, -10..10, 1..30),
     &                   integer npartons(1..30),
     &                   double precision xb(1..30),
     &                   double precision q2(1..30),
     &                   double precision xi(1..30),
     &                   double precision weight(1..11, 1..50),
     &                   integer irps(1..50),
     &                   integer ialphas(1..50),
     &                   integer ialphaem(1..50),
     &                   integer lognf(1..50)
     &                 )
\end{verbatim}

Here {\tt nr} stands for $R_n$, {\tt nl} stands for $L_n$, 
{\tt fvect(mu, iparticle, ir)} is the {\tt mu}$^{\mbox{th}}$ component
of the four-vector of the particle 
with label {\tt iparticle} ({\tt mu}=0 corresponds to the energy component)
in units of [GeV]
in the Breit frame 
for the phase space {\tt ir};
{\tt npartons(ir)} is the number of final-state partons,
{\tt q2(ir)} is the value of $Q^2$, and {\tt xi(ir)} is the momentum fraction
of the incident parton.
The particle labels {\tt iparticle} are given by
\begin{description}
\item[\quad{\tt iparticle=-8:}] proton remnant
\item[\quad{\tt iparticle=-7:}] incident proton
\item[\quad{\tt iparticle=-5:}] outgoing electron
\item[\quad{\tt iparticle=-4:}] incident electron
\item[\quad{\tt iparticle=-1:}] incident parton
\item[\quad{\tt iparticle=0..(npartons-1):}] outgoing partons
\end{description}

The array {\tt weight(j, i)} is a list of the weights for contribution
{\tt i} in units of [pb], 
{\tt irps(i)} gives the index of the phase space for this particular 
contribution,
{\tt ialphas(i)} and {\tt ialphaem(i)} are the powers of the strong 
and electromagnetic coupling constant, respectively, and {\tt lognf(i)}
is an index that specifies whether the weights have to be multiplied 
by a factor $\lambda$ consisting of a product of
a logarithm of a scale and/or a factor of $N_f$:
\begin{description}
\item[\quad{\tt lognf=0}:] $\lambda=1$
\item[\quad{\tt lognf=1}:] $\lambda=\ln\left(\mu_r^2/Q^2\right)$
\item[\quad{\tt lognf=2}:] $\lambda=N_f \ln\left(\mu_r^2/Q^2\right)$
\item[\quad{\tt lognf=3}:] $\lambda=\ln\left(\mu_f^2/Q^2\right)$
\item[\quad{\tt lognf=4}:] $\lambda=N_f \ln\left(\mu_f^2/Q^2\right)$
\end{description}
Here $\mu_r$ and $\mu_f$ are the renormalization and factorization scales, 
respectively.
The total flavour factor for contribution $i$ is given by 
\begin{equation}
F_{nij} = 
   \lambda \, 
   \alpha_s^{\mbox{\tt ialphas($i$)}} \,
   \alpha^{\mbox{\tt ialphem($i$)}} \,
   \rho_{ij},
\end{equation}
where 
the quantity $\rho_{ij}$ is a product of squares of quark charges $Q_\alpha$ 
in units of $e$ and parton densities.
In particular:
\begin{description}
\item \quad$\rho_{i1}
               = \sum\limits_{\alpha=1}^{N_f} Q_\alpha^2 \, f_\alpha$
\item \quad$\rho_{i2} 
               = \sum\limits_{\alpha=1}^{N_f} Q_\alpha^2 \,  
                 f_{\overline{\alpha}}$
\item \quad$\rho_{i3} 
               = \sum\limits_{\alpha=1}^{N_f} Q_\alpha^2 \, f_g$
\item \quad$\rho_{i4} = \rho_{i1}$
\item \quad$\rho_{i5} = \rho_{i2}$
\item \quad$\rho_{i6} = \rho_{i1}\,(N_f-1)$
\item \quad$\rho_{i7} = \rho_{i2}\,(N_f-1)$
\item \quad$\rho_{i8}
               = \sum\limits_{\alpha=1}^{N_f} f_\alpha \,
                 \sum\limits_{\beta=1,\, \beta \neq \alpha}^{N_f} Q_\beta^2$
\item \quad$\rho_{i9}
               = \sum\limits_{\alpha=1}^{N_f} f_{\overline{\alpha}} \,
                 \sum\limits_{\beta=1,\, \beta \neq \alpha}^{N_f} Q_\beta^2$
\item \quad$\rho_{i10}
               = \sum\limits_{\alpha=1}^{N_f} f_\alpha Q_\alpha \,
                 \sum\limits_{\beta=1,\, \beta \neq \alpha}^{N_f} Q_\beta$
\item \quad$\rho_{i11}
               = \sum\limits_{\alpha=1}^{N_f} f_{\overline{\alpha}} Q_\alpha \,
                 \sum\limits_{\beta=1,\, \beta \neq \alpha}^{N_f} Q_\beta$
\end{description}
The $f_\alpha$ are parton densities evaluated for 
momentum fractions $\mbox{\tt xi(irps($i$))}$ and factorization scale
$\mu_f$,
and $f_{\overline{\alpha}}$ stands for the parton density of the anti-flavour
of the flavour $\alpha$. The renormalization and factorization schemes are
the $\overline{\mbox{MS}}$ scheme. The correction terms for the 
DIS factorization 
scheme will be implemented in the near future.

We wish to note that the product of the weights, the flavour factors and the 
values of the observable is normalized in such a way that
the sum yields the expectation value in units of [pb]. No additional 
factor such as $1/N$, $N$ being the total number of generated events,
has to be applied
in Eq.~\ref{exval}.

Since phase spaces are used several times for different contributions,
it is a good strategy to first evaluate the observable(s) under consideration
for every phase space and to store the corresponding results.
Then there should be the loop over the various contributions (the second sum).
The innermost loop is the one over the flavour factors.

The Monte Carlo integration itself employs the program {\tt VEGAS}
\cite{27,28}. 
{\tt VEGAS} is an adaptive multi-dimensional integration routine.
Integrations proceed in two steps. 
The first step is an adaptation step in order
to set up a grid in the integration variables
which then steers the final integration step.
The adaptation step itself refines
the grid in a sequence of several iterations.
{\tt VEGAS} requires, as parameters, the number of Monte Carlo points 
to be used in the first and second step, respectively, 
and the number of iterations to refine the grid. 
In the framework of {\tt DISASTER++}, {\tt VEGAS} can be used in three different
ways: 
\begin{itemize}
\item As an adaptive integration routine.
The routine {\tt user2} should return a value. This value is handed over
to {\tt VEGAS} as 
the value of the integrand at the particular phase space point, 
and summed up. The final integral quoted by {\tt VEGAS}
is the sum of these weights
for the final integration.
This is the best choice if just one observable, 
for example a jet cross section, is to be evaluated.
\item As a routine that merely supplies random numbers for 
the events.
If the number of iterations is set to zero, then {\tt VEGAS} just performs
the final integration run. The user is then responsible for the correct
summation of the weights, and for the determination of the 
statistical error. It should be noted that, since all weights are
available individually in the user routine, an arbitrary number of 
observables can be evaluated in a single run. In particular, since the
dependence on the renormalization and factorization scales and on $N_f$
is fully explicit, the study of the scale dependence of observables
can be done in a very convenient way. For example, all plots from 
Ref.~\cite{22} 
have been obtained in a single run of {\tt DISASTER++}.
\item As a combination of the two preceeding alternatives. Here the adaptation 
steps are included. A ``typical'' infrared-safe observable, 
in the following called the {\it adaptation variable}, is evaluated, and
its value is returned to {\tt VEGAS}. This observable serves to optimize the
distribution of points over phase space. A convenient observable of this
kind is provided by {\tt DISASTER++} (see below).
The ``real'' observables under consideration are evaluated as in the 
second alternative in the final integration step.
\end{itemize}

\noindent
{\tt DISASTER++} is initialized by a call of 
the subroutine {\tt disaster\_ca()}. It is recommended 
to end a {\tt DISASTER++}
run by a call of the subroutine 
{\tt disaster\_cb()} in order to free
dynamically allocated memory.

\noindent
Parameters can be set or commands be executed by means of three routines:
\begin{description}
\item {\quad\tt disaster\_ci(str, i)} {\ }\\ 
   sets the integer parameter denoted by 
   the character string {\tt str} to the value {\tt i}
\item {\quad\tt disaster\_cd(str, d)} {\ }\\
   sets the double precision parameter denoted by 
   the character string {\tt str} to the value {\tt d}
\item {\quad\tt disaster\_cc(str)} {\ }\\ executes the command 
given by the character string {\tt str}
\end{description}
The following parameters are available (there are a few more to optimize the
generation of the phase space points; they will be documented in forthcoming
versions of this manual):
\begin{description}

\item[\quad{\makebox[3cm][l]{\tt ECM:}}]{\ }\\
   the centre-of-mass energy in [GeV]

\item[\quad{\tt LEPTON\_INTEGRATION}:]{\ }\\
   {\tt 1:} integration over $x_B$ and $y$

\item[\quad{\makebox[3cm][l]{\tt XBMIN:}}]{\ }\\
     minimum value of $x_B$

\item[\quad{\makebox[3cm][l]{\tt XBMAX:}}]{\ }\\
     maximum value of $x_B$

\item[\quad{\makebox[3cm][l]{\tt YMIN:}}]{\ }\\
     minimum value of $y$

\item[\quad{\makebox[3cm][l]{\tt YMAX:}}]{\ }\\
     maximum value of $y$

\item[\quad{\makebox[3cm][l]{\tt QMIN:}}]{\ }\\
     minimum value of $Q$

\item[\quad{\makebox[3cm][l]{\tt QMAX:}}]{\ }\\
     maximum value of $Q$

\item[\quad{\makebox[3cm][l]{\tt WMIN:}}]{\ }\\
     minimum value of $W$

\item[\quad{\makebox[3cm][l]{\tt WMAX:}}]{\ }\\
     maximum value of $W$

\item[\quad{\makebox[3cm][l]{\tt PROCESS\_INDEX:}}]{\ }\\
     {\tt 1:} leading order\\
     {\tt 2:} next-to-leading order
    
\item[\quad{\tt NUMBER\_OF\_FINAL\_STATE\_PARTONS\_IN\_BORN\_TERM:}]{\ }\\
     {\tt 1}, {\tt 2}, {\tt 3} for the process under consideration;\\
     {\tt 1:} (1+1)-jet-type observables (leading and next-to-leading order)\\
     {\tt 2:} (2+1)-jet-type observables (leading and next-to-leading order)\\
     {\tt 3:} (3+1)-jet-type observables (leading order only)
    
\item[\quad{\makebox[3cm][l]{\tt POINTS:}}]{\ }\\
     {\tt POINTS * (FACT\_PREP + FACT\_FINAL)} is the 
     number of generated points in the Monte Carlo integration

\item[\quad{\makebox[3cm][l]{\tt FACT\_PREP:}}]{\ }\\
     the number of points for the grid-definition run is given by
     {\tt FACT\_PREP * POINTS}

\item[\quad{\makebox[3cm][l]{\tt FACT\_FINAL:}}]{\ }\\
     the number of points for the final integration step is given by
     {\tt FACT\_FINAL * POINTS}

\item[\quad{\makebox[3cm][l]{\tt RUN\_MC:}}]{\ }\\
     to start the Monte-Carlo integration

\end{description}

\noindent
A convenient adaptation observable can be evaluated by a call of
the following function:
\begin{verbatim}
      double precision function disaster_cao(
     &                             integer ipdf_collection,
     &                             integer ipdf_parametrization,
     &                             integer ipdf_set,
     &                             integer ialphas_variant,
     &                             integer ialphas_order,
     &                             double precision dalphas_lambdaqcd4,
     &                             integer ialphaem_variant
     &                          )
\end{verbatim}
The arguments of the function call are:
\begin{description}

\item[\quad{\tt ipdf\_collection:}]{\ }\\
the collection of parton densities; \\
{\tt 1:} {\tt PDFLIB} \cite{29}

\item[\quad{\tt ipdf\_parametrization:}]{\ }\\
parametrization of parton densities (cf.\ {\tt PDFLIB})

\item[\quad{\tt ipdf\_set:}]{\ }\\
set of parton densities (cf.\ {\tt PDFLIB})

\item[\quad{\tt ialphas\_variant:}]{\ }\\
function which is used to evaluate the strong coupling constant;\\
{\tt 1:} running coupling $\alpha_s(Q^2)$ with 
flavour thresholds at the single heavy quark masses

\item[\quad{\tt ialphas\_order:}]{\ }\\
{\tt 1:} one-loop formula\\
{\tt 2:} two-loop formula\\
for the running strong
coupling constant

\item[\quad{\tt dalphas\_lambdaqcd4:}]{\ }\\
the QCD parameter $\Lambda_{\mbox{\scriptsize QCD}}^{(4)}$
for four flavours

\item[\quad{\tt ialphaem\_variant:}]{\ }\\
function which is used to evaluate the electromagnetic coupling constant;\\
{\tt 1:} fine structure constant \\
{\tt 2:} 1/137\\
(an implementation of the running electromagnetic 
coupling constant is in preparation)

\end{description}

\noindent
To simplify the calculation of the flavour factors, 
a {\tt DISASTER++} routine can be called which returns the
required coupling constants and the combinations of parton densities
and quark charges:
\begin{verbatim}
      subroutine disaster_cff(
     &              integer ipdf_collection,
     &              integer ipdf_parametrization,
     &              integer ipdf_set,
     &              integer ialphas_variant,
     &              integer ialphas_order,
     &              double precision dalphas_lambdaqcd4,
     &              integer ialphaem_variant,
     &              integer nf,
     &              double precision ffactin(4),
     &              double precision ffactout(13)
     &           )
\end{verbatim}
The arguments of the function call are the same as in the case of the
routine {\tt disaster\_cao} (see above), except for the following:
\begin{description}

\item[\quad{\tt nf:}]{\ }\\
the number of flavours $N_f$

\item[\quad{\tt ffactin:}]{\ }\\
input parameters;
\begin{description}
\item {\tt ffactin(1):} the momentum fraction variable $\xi$
\item {\tt ffactin(2):} the factorization scale in [GeV] 
(i.e.\ the scale argument of the parton densities)
\item {\tt ffactin(3):} the renormalization scale in [GeV] 
(i.e.\ the scale argument of the running strong coupling constant)
\item {\tt ffactin(4):} the scale argument of the running electromagnetic
coupling constant
\end{description}

\item[\quad{\tt ffactout:}]{\ }\\
output parameters;
\begin{description}
\item {\tt ffactout(1..11):} the quantities $\rho_{i1}$ \ldots
$\rho_{i11}$,
\item {\tt ffactout(12):} the running strong coupling constant
\item {\tt ffactout(13):} the electromagnetic
coupling constant
\end{description}

\end{description}

It is strongly recommended to use this routine, since it uses 
a cache that stores a few of the most recent values temporarily, such that
the sums $\rho_{ij}$ and the parton densities do not have to be reevaluated.
This routine is supplied for the convenience of the user. The weights
and events generated by {\tt DISASTER++} do not depend on this routine.

The description of the program structure just given may sound
complicated. It is actually quite simple to use the program; an example 
for the calculation of the (2+1)-jet cross section for the JADE algorithm
in the E-scheme is given in the files {\tt disaster\_f.f}
and {\tt clust.f}, as described in Section~\ref{installation}.

\section{Program Installation}
\label{installation}

\begin{description}
\item[Source code:]
The source code of the class library is available on the World Wide Web:
\begin{verbatim}
   http://wwwcn.cern.ch/~graudenz/disaster.html
\end{verbatim}

\item[Files:]
The package consists of a number of files. To facilitate the installation,
and to enable the {\tt C++} compiler to perform certain optimizations,
the complete {\tt C++} part of the program is provided as one file
{\tt onefile\_n.cc} (the individual files are available on request). 
An example for the {\tt FORTRAN} interface is
given in the file {\tt disaster\_f.f} (calculation of the (2+1) jet 
cross section for the JADE algorithm in the E-scheme), 
together with a simple cluster
routine in the file {\tt clust.f}. 
The number of Monte Carlo events in the example is set to 
a tiny number (100) in order to terminate the program after a few seconds.
Realistic values for the parameter {\tt POINTS} are of the order of 
$10^6$.
An example ``make file'' is given in {\tt makedisaster}. 
\item[Mixed Language Programming:]
{\tt DISASTER++} is mainly writen in the {\tt C++} programming language.
The reason for the choice of this language are twofold:
Object-oriented programming allows for programs that are easily 
maintained and extended\footnote{It could even be said that object-oriented
programming is a kind of applied ontology: the central categories of this 
approach are given by {\it objects} and {\it methods} that define their
relationships.
}, and in high-energy physics there is a trend in the experimental 
domain for a transition from {\tt FORTRAN} to {\tt C++}.
Although the goal has been to write a self-contained {\tt C++}
package, 
a few parts of the program are still coded in 
{\tt FORTRAN}. Moreover, the standard parton density parametrizations
are only 
available as {\tt FORTRAN} libraries. This means that the {\tt DISASTER++}
package cannot be run as a stand-alone {\tt C++} program. In most cases,
users may wish to interface the program to their existing {\tt FORTRAN}
routines. An elegant and machine-independent 
way for {\it mixed language programming} for the case
of {\tt C}, {\tt C++} and {\tt FORTRAN} is supported by the 
{\tt cfortran.h} package described in Ref.~\cite{30}. 
For every {\tt FORTRAN} routine to be called by a {\tt C++} method, 
an {\tt extern "C"} routine has to be defined as an interface, 
and vice versa. The explicit calls are then generated by means of macros 
from {\tt cfortran.h}. The most convenient way is, after compilation, 
to link the {\tt FORTRAN} and {\tt C++} parts via the standard
\begin{verbatim}
   f77 -o disaster onefile_n.o ...
\end{verbatim}
command\footnote{The procedure is described here for the {\tt UNIX}
operating system.}, 
such that the {\tt FORTRAN} part supplies the entry point.
The required {\tt C++} libraries have to be stated explicitly
via the {\tt -L} and {\tt -l} options. The library paths can be obtained
by compiling and linking a trivial program {\tt hw.cc} of the type
\begin{verbatim}
   #include <stdio.h>
   main() { printf("Hello world!\n"); }
\end{verbatim}
with
\begin{verbatim}
   gcc -v hw.cc
\end{verbatim}
(for the {\tt GNU C++} compiler). 
An example for the required libraries can be found in the 
prototype ``make file'' {\tt makedisaster}. Some machine-specific information
is mentioned in the manual of {\tt cfortran.h}.

In the {\tt DISASTER++} package, the explicit {\tt FORTRAN} interface, 
as described in Section~\ref{structure},
is already provided. Thus
from the outside the {\tt C++}
kernel is transparent and hidden behind {\tt FORTRAN} subroutines
and functions.

\item[Template instantiation:]
In {\tt DISASTER++}, heavy use is made of {\it templates}. At present, there
is not yet a universally accepted scheme for template instantiations.
The solution adopted here is the explicit instantiation
of all templates. This requires
that the compiler itself does not instantiate templates automatically.
This is achieved for the {\tt GNU} compiler by means of the switch
\begin{verbatim}
   -fno-external-templates
\end{verbatim}

\item[Output files:]
There is a small problem with the output from the {\tt C++} and {\tt FORTRAN}
parts of {\tt DISASTER++}. It seems to be the case that generally {\tt C++} 
({\tt FILE* stdout} and {\tt ostream cout}) 
and {\tt FORTRAN} ({\tt UNIT=6}) keep different
file buffers. This is no problem when the output is written to a terminal, 
since then the file buffers are immediately flushed
after each line-feed character. When writing to 
a file (as is usually the case for batch jobs), the file buffers are not 
immediately flushed, and this leads to the problem that the output 
on the file is mixed in non-chronological order. This problem will be solved
by the introduction of a particular stream class which hands over the output
to a {\tt FORTRAN} routine.

\item[Miscellaneous:]
{\tt DISASTER++} employs the {\tt ANSI C} {\tt signal} facility to 
catch interrupts caused by floating point arithmetic. 
If the signal {\tt SIGFPE} is raised, a flag in {\tt DISASTER++} is set, 
which eventually leads to the requirement that the event has to be 
dropped (via a call of {\tt user1(7)}). Similarly a non-zero value of
the variable {\tt errno} of the {\tt ANSI C errno} facility
is treated. The signal handler is also active 
when the user routine is executed, which leads to the effect that
in the case of a floating point exception the program does not crash, but
continues under the assumption that the event has been dropped.
Forthcoming version of {\tt DISASTER++} will make a flag available
that can be used to access the status of the signal handler in 
the user routines. 
Moreover, it is checked whether the weight returned to {\tt DISASTER++} via
{\tt user2} fulfills the criterion for {\tt IEEE} {\tt NaN} (``not a number''). 
If this is the case, it is also requested that the event be dropped.

\end{description}

\section{Comparison of Available Programs}
\label{comparison}

In this section, we compare the three available programs {\tt MEPJET}
(Version 2.0)\footnote{
For the very-high statistics runs the default random number
generator (generating a Sobol sequence of pseudo-random numbers) 
of {\tt MEPJET} ran out of random numbers. We therefore had to modify the
program such that it uses another generator which is also part of
the {\tt MEPJET} package. --- The crossing functions for the ``artificial''
parton densities have been obtained by means of a modification of the program
{\tt make\_str\_pdf1.f}.
}, 
{\tt DISENT} (Version 0.1)\footnote{
An earlier version of this paper \cite{24}
reported results of a comparison 
with {\tt DISENT} Version 0.0. We found large discrepancies for 
some choices of the parton density parametrization. In the meantime,
an error in {\tt DISENT} has been fixed, and the results 
of {\tt DISENT} and {\tt DISASTER++} are now
in good agreement, see below.
} and {\tt DISASTER++}
(Version 1.0) numerically for various bins of 
$x_B$ and $y$ as defined in Table~\ref{tab1}, 
and for various choices of the parton density parametrizations.

\begin{table}[htb]
\begin{center}
\begin{tabular}[h]{|c|c|c|c|}
\cline{2-4}
  \multicolumn{1}{c|}{\rule[-2.5mm]{0mm}{8mm}}
  & \makebox[4.1cm]{$0.01 < y < 0.03 $}  
  & \makebox[4.1cm]{$0.03 < y < 0.1 $}  
  & \makebox[4.1cm]{$0.1  < y < 0.3 $}  
\\ \hline
$0.005 < x_B < 0.01$ \rule[-2.5mm]{0mm}{8mm}
  & \makebox[1.2cm]{Bin 1}($Q^2 > 4.6\,\mbox{GeV}^2$)
  & \makebox[1.2cm]{Bin 2}($Q^2 > 13.5\,\mbox{GeV}^2$)
  & \makebox[1.2cm]{Bin 3}($Q^2 > 45.0\,\mbox{GeV}^2$)
\\ \hline
$0.05 < x_B < 0.1$ \rule[-2.5mm]{0mm}{8mm}
  & \makebox[1.2cm]{Bin 4} ($Q^2 > 45\,\mbox{GeV}^2$)
  & \makebox[1.2cm]{Bin 5} ($Q^2 > 135\,\mbox{GeV}^2$)
  & \makebox[1.2cm]{Bin 6} ($Q^2 > 450\,\mbox{GeV}^2$)
\\ \hline
$0.2 < x_B < 0.4$ \rule[-2.5mm]{0mm}{8mm}
  & \makebox[1.2cm]{Bin 7} ($Q^2 > 180\,\mbox{GeV}^2$)
  & \makebox[1.2cm]{Bin 8} ($Q^2 > 540\,\mbox{GeV}^2$)
  & \makebox[1.2cm]{Bin 9} ($Q^2 > 1800\,\mbox{GeV}^2$)
\\ \hline
\end{tabular}
\end{center}
\caption[tab1]
{
\label{tab1}
{\it
Bins in $x_B$ and $y$. The values in parentheses give the resulting 
lower bounds on $Q^2$.
}
}
\end{table}

The centre-of-mass energy is set to 300\,GeV. To facilitate the
comparison, the strong coupling constant is set to a fixed value of
$\alpha_s=0.1$, 
and the number of flavours is set to $N_f=5$, even below the bottom 
threshold ($N_f=5$ is hard-wired into {\tt MEPJET}). 
The electromagnetic coupling constant 
is chosen to be $\alpha=1/137$ (the value
is hard-wired into {\tt DISENT}, but this could be changed trivially, 
in principle). The factorization- and renormalization schemes of the
hard scattering cross sections are $\overline{\mbox{MS}}$, and the
factorization and renormalization scales $\mu_f$ and $\mu_r$, respectively, 
are set to $Q$.

The quantity under consideration is the (2+1)-jet cross section, 
shown in Tables~2--8 in Appendix~\ref{capp}.
For simplicity we consider the modified JADE clustering scheme
with resolution criterion $S_{ij} <> c W^2$ and the E~recombination scheme, 
where
$S_{ij} = (p_i + p_j)^2$, $W$ is the total hadronic energy,
and $c=0.02$ is the jet resolution parameter.
We require, in the laboratory frame ($E_e=27.439$\,GeV, 
$E_P=820$\,GeV), a minimum transverse momentum of 1\,GeV and a pseudo-rapidity
of $-3.5<\eta<3.5$ for all jets\footnote{These cuts in $p_T$ and $\eta$ 
are employed in order to facilitate event generation with {\tt MEPJET}; 
the phase space generator implemented in that program is reminiscent of a 
generator for pp~collider physics where $p_T$ and $\eta$ cuts
in the laboratory frame are
a standard experimental procedure. It is thus complicated to generate 
events with {\tt MEPJET} 
in the full phase space of the laboratory system, as usually required 
for eP scattering, where ``natural'' cuts in transverse momentum and
pseudo-rapidity would be performed in the hadronic centre-of-mass frame
or in the Breit frame.}.

The parton density parametrizations employed in the comparison are:
\begin{description}
\item[{\makebox[1cm][l]{(a)}}] 
           \makebox[6cm][l]{the MRSD$_-^\prime$ parton densities 
                            \cite{31}} (Table 2),
\item[{\makebox[1cm][l]{(b)}}] 
           \makebox[3cm][l]{$q(\xi)=(1-\xi)^5$,}\makebox[3cm][l]{$g(\xi)=0$}
           (Table 3),
\item[{\makebox[1cm][l]{(c)}}] 
           \makebox[3cm][l]{$q(\xi)=0$,}\makebox[3cm][l]{$g(\xi)=(1-\xi)^5$} 
           (Table 4),
\item[{\makebox[1cm][l]{(d)}}] 
           \makebox[3cm][l]{$q(\xi)=(1-\xi)^2$,}\makebox[3cm][l]{$g(\xi)=0$}
           (Table 5),
\item[{\makebox[1cm][l]{(e)}}] 
           \makebox[3cm][l]{$q(\xi)=0$,}\makebox[3cm][l]{$g(\xi)=(1-\xi)^2$} 
           (Table 6),
\item[{\makebox[1cm][l]{(f)}}] 
           \makebox[3cm][l]{$q(\xi)=(1-\xi)$,}\makebox[3cm][l]{$g(\xi)=0$}
           (Table 7),
\item[{\makebox[1cm][l]{(g)}}] 
           \makebox[3cm][l]{$q(\xi)=0$,}\makebox[3cm][l]{$g(\xi)=(1-\xi)$} 
           (Table 8).
\end{description}
Here $q(\xi)$ generically stands for valence and sea distributions\footnote{
This means that $u_v(\xi)$, $d_v(\xi)$, $u_s(\xi)$, $d_s(\xi)$, 
$s_s(\xi)$, $c_s(\xi)$, 
$b_s(\xi)$ have been set to $q(\xi)$.
}, 
and $g(\xi)$ is the gluon distribution.
We wish to point out that the comparison involving the ``artificial''
parton densities is not just of academic interest. On the contrary, 
for the extraction 
of, for instance, the gluon density
from jet data 
it is convenient
to replace the parton densities by simple functions 
with special properties (such as powers
of the momentum fraction variable $\xi$ or functions of an orthonormal 
basis system), 
in order to achieve a fast fit. These functions usually do not have
the shape of physical parton densities, in particular they do not
have to fall off rapidly for $\xi\rightarrow 1$.
Moreover, next-to-leading-order calculations yield unique and well-defined
results for the hard scattering cross sections to be convoluted with 
observables and parton densities. We employ the ``artificial''
parton densities also in order to have a stricter test of the 
hard scattering cross sections.

The leading-order results of all three programs are in excellent agreement. 
The next-to-leading-order results of {\tt DISASTER++} and {\tt DISENT} are
in good agreement within about two to (sometimes) 
three standard deviations\footnote{
We wish to note that the error estimates quoted by the programs are usually not
rigorous estimates because of the non-Gaussian distribution of the
Monte-Carlo weights. Therefore, in principle, it is not possible to 
infer probabilities for the consistency of data samples produced by two
programs based on these estimates. 
A more precise, but in general unfeasible way to obtain an estimate of the
Monte Carlo error would be to run the programs a number of times 
with different random number seeds and to analyze the spread of the 
quoted results around their central value.
Such a study has recently been done by M.~Seymour
for {\tt DISENT} with the result that the 
individual error estimates are quite reliable \cite{32}.
}
of the larger of the two errors quoted by the two programs. An exception
is bin~7 for $g(\xi) = (1-\xi)^2$. A run of {\tt DISENT} with higher statistics 
yields a value of $0.1836 \pm 0.0025$, which is within two standard deviations
of the {\tt DISASTER++} result, indicating that there was indeed a statistical
fluctuation in the original result.

The comparison of the next-to-leading-order results 
of {\tt MEPJET} and {\tt DISASTER++} requires a more detailed discussion:
\begin{itemize}

\item For the MRSD$_-^\prime$ parton densities, the results for
bins 3--9 are
compatible within about two standard deviations of the statistical error
of the Monte-Carlo integrations.
The results for bins~1 and~2 differ considerably. 
Runs with a smaller value
of the internal {\tt MEPJET} cut-off variable~$s$, which is set by default
to $s=0.1\,$GeV$^2$, yield
the following results for bin 1:
$580.6 \pm 6.7$\,pb ($s=0.01\,$GeV$^2$), 
$564.8 \pm 10.5$\,pb ($s=0.001\,$GeV$^2$) and
$575.4 \pm 13.0$\,pb ($s=0.0001\,$GeV$^2$).
The statistical error is increased for decreased~$s$ because the integration
volume
of the (3+1) parton contributions is extended into the singular domain.
Because of the increased statistical error, we also performed a 
high-statistics runs with $\sim 4\cdot10^9$ (!) Monte Carlo events
of {\tt MEPJET} 
for this bin. 
For $s=0.001\,$GeV$^2$ we obtain 
$576.3 \pm 6.7$\,pb
and for $s=0.0001\,$GeV$^2$
the result is
$583.2 \pm 7.4$\,pb.
These values from {\tt MEPJET} 
are compatible with the {\tt DISASTER++} and {\tt DISENT} results\footnote{
These results underline that, for the phase space slicing method, results
generally have to be validated {\it ex post} by a cross-check with a 
smaller technical cut~$s$ and much higher statistics. It may be argued that
there are jet algorithms (the $k_T$~algorithm, for example)
which show a better convergence for $s\rightarrow 0$.
However, the point here is that one does not know in advance whether this
is the case for the observable under consideration. --- In Ref.~\cite{23}
we find the statement that $s$-independence in {\tt MEPJET} is achieved for 
$s=0.1\,$GeV$^2$. Our study shows that this is generally not the case, 
and that extremely small values of~$s$, possibly of the order of
$s=0.0001\,$GeV$^2$, might be necessary.
}.
\item For the parton density parametrization (b) (quarks only, with a steeply
falling distribution $q(\xi)$ for $\xi \rightarrow 1$), 
{\tt DISASTER++} and {\tt MEPJET}
are in good agreement.

\item The results for parametrization (c) (steeply falling
gluon parametrization)
are in good agreement, except for bin 1.

\item For parametrization (d), 
{\tt DISASTER++} and {\tt MEPJET} are in agreement except for bins 1 and 4.
Runs with a smaller value
of the {\tt MEPJET} cut-off variable~$s$
yield
the following results for bin 1:
$59.6 \pm 1.8$\,pb ($s=0.01\,$GeV$^2$), 
$56.7 \pm 5.8$\,pb ($s=0.001\,$GeV$^2$) and
$54.9 \pm 10.4$\,pb ($s=0.0001\,$GeV$^2$).
A high-statistics run ($\sim 4\cdot10^9$ events) of {\tt MEPJET} 
for bin 1 with $s=0.0001\,$GeV$^2$ gives the 
cross section $60.0 \pm 1.9$\,pb.
Contrary to the observation in case (a), for small~$s$ 
we do not get agreement of 
the {\tt MEPJET} result with the {\tt DISASTER++} / {\tt DISENT} result
of about $48$--$49$\,pb.

\item The {\tt MEPJET} results for parametrization (e) 
($g(\xi) = (1-\xi)^2$)
deviate considerably from the {\tt DISASTER++}
results in bins~1, 2, 4 and 7.

\item For parametrization (f),
{\tt DISASTER++} and {\tt MEPJET} are incompatible
for bins 1, 2, 4, 6 and 7.

\item For parametrization (g), 
{\tt MEPJET} and {\tt DISASTER++} are compatible in bins
3, 5, 8 and 9 only.
A high-statistics run ($\sim 4\cdot10^9$ events) of {\tt MEPJET} 
for bin 4 with $s=0.0001\,$GeV$^2$ yields the 
cross section $1.29 \pm 0.02$\,pb.
This value is different from the result for $s=0.1\,$GeV$^2$, 
but still inconsistent
with the {\tt DISASTER++} / {\tt DISENT} result of about $0.69$\,pb.

\end{itemize}

The overall picture is thus: Out of the three programs, {\tt DISASTER++} 
and {\tt DISENT} (Version 0.1) are in good agreement within about
two, sometimes three standard deviations of the quoted integration errors, 
both for ``physical'' and ``artificial'' parton densities. This agreement
is very encouraging, but not yet perfect, and much more detailed studies
involving different sets of observables and differential distributions
are required. For the two programs, a direct comparison of the
``jet structure functions'' should also be feasible.

For several bins, in particular for the ``artificial'' parton distribution 
functions, the {\tt MEPJET}
results for the default setting of the 
internal parameters deviate considerably from the {\tt DISASTER++}
and {\tt DISENT} results. 
For one particular bin studied in more detail for
the MRSD$_-^\prime$ parton densities,
the
discrepancy disappears in the case of an extremely small internal technical 
cut~$s$ of {\tt MEPJET}, for a substantial increase of the
number of generated events to obtain a meaningful Monte Carlo error. 
A few {\tt MEPJET} results employing ``artificial'' 
parton densities have been studied in more detail. We observed that 
in these cases a reduction of the~$s$ parameter does not lead to an
improvement of the situation. For lack of computer time, we could not study 
all bins with a smaller $s$~cut. The overall situation 
is thus still inconclusive and unclear. An independent cross check of the
{\tt MEPJET} results, in particular of those using the 
implementation of the crossing functions for the ``artificial'' parton 
densities, is highly desirable.

\section{Miscellaneous}
\begin{itemize}

\item If you intend to install and use {\tt DISASTER++}, please send me 
a short e-mail message, and I will put your name on a mailing list
so that I can inform you when there is a new version of the package.

\item Suggestions for improvements and bug reports are welcome.

\item In case that there are problems with the installation of the program, 
please send me an e-mail message.

\end{itemize}

\section{Summary}

We have presented the {\tt C++} class library 
{\tt DISASTER++} for the calculation 
of (1+1) and (2+1)-jet type observables in deeply inelastic scattering.
The program is based on the subtraction formalism and thus does not require
a technical cut-off for the separation of the infrared-singular from the
infrared-finite phase-space regions. 
A {\tt FORTRAN} interface to the {\tt C++} class library is available.
{\tt DISASTER++} is actually intended to be a general object-oriented
framework for next-to-leading order QCD calculations. In particular, 
the subtraction formalism is implemented in a very general way.

We have performed a comparison of the three available programs
{\tt MEPJET}, {\tt DISENT} and {\tt DISASTER++}
over a wide range of the parameters for the lepton phase space.
We find good agreement of {\tt DISASTER++} and the Catani-Seymour
program {\tt DISENT} (Version 0.1).
The comparison of {\tt DISASTER++} and the Mirkes-Zeppenfeld program
{\tt MEPJET} (for the {\tt MEPJET} 
default parameters) leads to several
discrepancies, both for physical and for ``artificial'' parton densities.
For the MRSD$_-^\prime$ parton densities a 
reduction of the internal {\tt MEPJET} phase-space slicing cut-off 
variable~$s$, the number of Monte Carlo events kept fixed, leads to a certain 
improvement of the central values of the results, 
accompanied by a substantially increased statistical error and fluctuating
central values. A considerable increase of the number of generated events
(up to of the order of several billion events) 
eventually leads to an agreement of the {\tt MEPJET} results with the
{\tt DISASTER++} / {\tt DISENT} results for a particular bin of the lepton 
variables which has been studied in detail.
For ``artificial'' parton densities and a selected set of bins of
the lepton variables, a reduction of the internal cut~$s$
does not resolve the discrepancies.
Other bins are not considered
for the lack of computer time for very-high statistics runs.
It should be stressed that the present study is still limited in scope.
An independent cross check of the {\tt MEPJET} results for the ``artificial''
parton densities has to be done until a firm conclusion can be reached.
Moreover, 
the study has to be repeated for a wider range of observables and much higher
Monte Carlo statistics. The $s$~dependence of the {\tt MEPJET} results
should also be studied in more detail.

Compared to the other two programs {\tt MEPJET} and {\tt DISENT},
{\tt DISASTER++} makes the full $N_f$ dependence and the dependence
on the renormalization and factorization scales available in the 
user routine. This is required for consistent studies of effects
such as the scale dependence when the bottom threshold is crossed.

\section{Acknowledgements}
I wish to thank M.~Seymour for sending me the numerical results for the new 
{\tt DISENT} version. D.~Zeppenfeld made a few cross
checks of the results for the MRSD$_-^\prime$ parton densities.
J.~Collins has provided me with the {\tt FORTRAN} 
routine to test the {\tt IEEE NaN} condition.
I am also grateful to Th.~Hadig for a few comments on the first version 
of this paper, and for suggestions for improvements of the program.

\clearpage

\begin{appendix}

\section{Numerical Results}
\label{capp}

This appendix contains the numerical results which are discussed in 
Section~\ref{comparison}. The entries in the tables are the (2+1)-jet 
cross sections
in units of [pb].

\begin{center}
\begin{tabular}[h]{|c|c|c|c|c|c|c|}
\cline{2-7}
   \multicolumn{1}{c|}{\rule[-2.5mm]{0mm}{8mm}}
 & \multicolumn{3}{|c|}{leading order}
 & \multicolumn{3}{|c|}{next-to-leading order}
\\ \hline
  bin\rule[-2.5mm]{0mm}{8mm}
  & \makebox[2.2cm]{\tt DISASTER++}  
  & \makebox[2.2cm]{\tt MEPJET}  
  & \makebox[2.2cm]{\tt DISENT}  
  & \makebox[2.2cm]{\tt DISASTER++}  
  & \makebox[2.2cm]{\tt MEPJET}  
  & \makebox[2.2cm]{\tt DISENT}  
\\ \hline\hline
1\rule[-2.5mm]{0mm}{8mm}
  & \pmdg{402.1}{1.13}
  & \pmdg{399.9}{0.53}
  & \pmdg{399.6}{1.1}
  & \pmdg{585.0}{2.6}
  & \pmdg{564.1}{1.9}
  & \pmdg{578.4}{7.1}
\\ \hline
2\rule[-2.5mm]{0mm}{8mm}
  & \pmdg{207.6}{0.59}
  & \pmdg{207.5}{0.34}
  & \pmdg{207.4}{0.15}
  & \pmdg{364.8}{1.5}
  & \pmdg{347.3}{2.4}
  & \pmdg{361.1}{3.5}
\\ \hline
3\rule[-2.5mm]{0mm}{8mm}
  & \pmdg{60.0}{0.16}
  & \pmdg{59.9}{0.14}
  & \pmdg{59.9}{0.15}
  & \pmdg{119.1}{1.71}
  & \pmdg{118.0}{1.05}
  & \pmdg{120.1}{0.94}
\\ \hline
4\rule[-2.5mm]{0mm}{8mm}
  & \pmdg{82.9}{0.16}
  & \pmdg{82.9}{0.10}
  & \pmdg{82.6}{0.21}
  & \pmdg{98.1}{1.11}
  & \pmdg{95.1}{0.61}
  & \pmdg{95.4}{0.87}
\\ \hline
5\rule[-2.5mm]{0mm}{8mm}
  & \pmdg{42.9}{0.08}
  & \pmdg{42.9}{0.06}
  & \pmdg{42.6}{0.28}
  & \pmdg{55.3}{0.46}
  & \pmdg{54.4}{0.49}
  & \pmdg{54.9}{0.40}
\\ \hline
6\rule[-2.5mm]{0mm}{8mm}
  & \pmdg{11.9}{0.02}
  & \pmdg{11.9}{0.02}
  & \pmdg{11.9}{0.08}
  & \pmdg{17.5}{0.06}
  & \pmdg{16.8}{0.22}
  & \pmdg{17.3}{0.13}
\\ \hline
7\rule[-2.5mm]{0mm}{8mm}
  & \pmdg{9.60}{0.03}
  & \pmdg{9.58}{0.01}
  & \pmdg{9.59}{0.04}
  & \pmdg{12.1}{0.50}
  & \pmdg{12.7}{0.07}
  & \pmdg{12.3}{0.15}
\\ \hline
8\rule[-2.5mm]{0mm}{8mm}
  & \pmdg{6.24}{0.01}
  & \pmdg{6.23}{0.01}
  & \pmdg{6.24}{0.02}
  & \pmdg{8.61}{0.12}
  & \pmdg{8.55}{0.15}
  & \pmdg{8.52}{0.08}
\\ \hline
9\rule[-2.5mm]{0mm}{8mm}
  & \pmdg{1.78}{0.003}
  & \pmdg{1.78}{0.003}
  & \pmdg{1.78}{0.06}
  & \pmdg{2.65}{0.03}
  & \pmdg{2.57}{0.06}
  & \pmdg{2.63}{0.02}
\\ \hline
\end{tabular}

\vspace{0.5cm}
Table 2: {\it
Comparison for MRSD$_-^{\,\prime}$ parton densities.
}
\end{center}

\clearpage

\begin{center}
\begin{tabular}[h]{|c|c|c|c|c|c|c|}
\cline{2-7}
   \multicolumn{1}{c|}{\rule[-2.5mm]{0mm}{8mm}}
 & \multicolumn{3}{|c|}{leading order}
 & \multicolumn{3}{|c|}{next-to-leading order}
\\ \hline
  bin\rule[-2.5mm]{0mm}{8mm}
  & \makebox[2.2cm]{\tt DISASTER++}  
  & \makebox[2.2cm]{\tt MEPJET}  
  & \makebox[2.2cm]{\tt DISENT}  
  & \makebox[2.2cm]{\tt DISASTER++}  
  & \makebox[2.2cm]{\tt MEPJET}  
  & \makebox[2.2cm]{\tt DISENT}  
\\ \hline\hline
1\rule[-2.5mm]{0mm}{8mm}
  & \pmdg{36.2}{0.09}
  & \pmdg{36.3}{0.05}
  & \pmdg{36.3}{0.12}
  & \pmdg{39.1}{0.33}
  & \pmdg{40.9}{0.89}
  & \pmdg{38.2}{0.53}
\\ \hline
2\rule[-2.5mm]{0mm}{8mm}
  & \pmdg{17.8}{0.04}
  & \pmdg{17.8}{0.03}
  & \pmdg{17.7}{0.05}
  & \pmdg{23.2}{0.37}
  & \pmdg{22.7}{0.41}
  & \pmdg{22.6}{0.22}
\\ \hline
3\rule[-2.5mm]{0mm}{8mm}
  & \pmdg{5.21}{0.01}
  & \pmdg{5.21}{0.01}
  & \pmdg{5.21}{0.02}
  & \pmdg{8.24}{0.22}
  & \pmdg{7.86}{0.12}
  & \pmdg{8.14}{0.06}
\\ \hline
4\rule[-2.5mm]{0mm}{8mm}
  & \pmdg{27.3}{0.06}
  & \pmdg{27.3}{0.03}
  & \pmdg{27.2}{0.09}
  & \pmdg{28.0}{0.52}
  & \pmdg{29.2}{0.18}
  & \pmdg{30.0}{0.21}
\\ \hline
5\rule[-2.5mm]{0mm}{8mm}
  & \pmdg{14.8}{0.03}
  & \pmdg{14.8}{0.02}
  & \pmdg{14.7}{0.04}
  & \pmdg{17.4}{0.29}
  & \pmdg{16.9}{0.10}
  & \pmdg{17.0}{0.11}
\\ \hline
6\rule[-2.5mm]{0mm}{8mm}
  & \pmdg{4.33}{0.008}
  & \pmdg{4.32}{0.006}
  & \pmdg{4.31}{0.01}
  & \pmdg{5.62}{0.10}
  & \pmdg{5.44}{0.05}
  & \pmdg{5.54}{0.03}
\\ \hline
7\rule[-2.5mm]{0mm}{8mm}
  & \pmdg{6.38}{0.02}
  & \pmdg{6.37}{0.01}
  & \pmdg{6.38}{0.03}
  & \pmdg{8.49}{0.17}
  & \pmdg{8.59}{0.10}
  & \pmdg{8.37}{0.11}
\\ \hline
8\rule[-2.5mm]{0mm}{8mm}
  & \pmdg{4.44}{0.01}
  & \pmdg{4.43}{0.007}
  & \pmdg{4.44}{0.02}
  & \pmdg{6.11}{0.08}
  & \pmdg{6.05}{0.07}
  & \pmdg{6.07}{0.06}
\\ \hline
9\rule[-2.5mm]{0mm}{8mm}
  & \pmdg{1.36}{0.002}
  & \pmdg{1.36}{0.002}
  & \pmdg{1.36}{0.05}
  & \pmdg{2.02}{0.02}
  & \pmdg{2.00}{0.05}
  & \pmdg{2.01}{0.01}
\\ \hline
\end{tabular}

\vspace{0.5cm}
Table 3: {\it
Comparison for $q(\xi) = (1-\xi)^5$
}
\end{center}

\begin{center}
\begin{tabular}[h]{|c|c|c|c|c|c|c|}
\cline{2-7}
   \multicolumn{1}{c|}{\rule[-2.5mm]{0mm}{8mm}}
 & \multicolumn{3}{|c|}{leading order}
 & \multicolumn{3}{|c|}{next-to-leading order}
\\ \hline
  bin\rule[-2.5mm]{0mm}{8mm}
  & \makebox[2.2cm]{\tt DISASTER++}  
  & \makebox[2.2cm]{\tt MEPJET}  
  & \makebox[2.2cm]{\tt DISENT}  
  & \makebox[2.2cm]{\tt DISASTER++}  
  & \makebox[2.2cm]{\tt MEPJET}  
  & \makebox[2.2cm]{\tt DISENT}  
\\ \hline\hline
1\rule[-2.5mm]{0mm}{8mm}
  & \pmdg{4.89}{0.017}
  & \pmdg{4.89}{0.007}
  & \pmdg{4.87}{0.01}
  & \pmdg{5.38}{0.07}
  & \pmdg{6.03}{0.06}
  & \pmdg{5.22}{0.13}
\\ \hline
2\rule[-2.5mm]{0mm}{8mm}
  & \pmdg{2.66}{0.009}
  & \pmdg{2.66}{0.007}
  & \pmdg{2.65}{0.007}
  & \pmdg{3.67}{0.08}
  & \pmdg{3.66}{0.04}
  & \pmdg{3.58}{0.05}
\\ \hline
3\rule[-2.5mm]{0mm}{8mm}
  & \pmdg{0.825}{0.003}
  & \pmdg{0.826}{0.002}
  & \pmdg{0.826}{0.002}
  & \pmdg{1.44}{0.07}
  & \pmdg{1.37}{0.03}
  & \pmdg{1.39}{0.02}
\\ \hline
4\rule[-2.5mm]{0mm}{8mm}
  & \pmdg{1.60}{0.005}
  & \pmdg{1.60}{0.003}
  & \pmdg{1.60}{0.003}
  & \pmdg{1.20}{0.05}
  & \pmdg{1.30}{0.01}
  & \pmdg{1.12}{0.04}
\\ \hline
5\rule[-2.5mm]{0mm}{8mm}
  & \pmdg{0.904}{0.003}
  & \pmdg{0.900}{0.001}
  & \pmdg{0.899}{0.002}
  & \pmdg{0.833}{0.027}
  & \pmdg{0.801}{0.008}
  & \pmdg{0.764}{0.019}
\\ \hline
6\rule[-2.5mm]{0mm}{8mm}
  & \pmdg{0.279}{0.001}
  & \pmdg{0.278}{0.001}
  & \pmdg{0.278}{0.001}
  & \pmdg{0.314}{0.007}
  & \pmdg{0.287}{0.004}
  & \pmdg{0.299}{0.006}
\\ \hline
7\rule[-2.5mm]{0mm}{8mm}
  & \pmdg{0.130}{0.001}
  & \pmdg{0.131}{0.001}
  & \pmdg{0.130}{0.001}
  & \pmdg{0.119}{0.005}
  & \pmdg{0.118}{0.002}
  & \pmdg{0.110}{0.006}
\\ \hline
8\rule[-2.5mm]{0mm}{8mm}
  & \pmdg{0.0981}{0.001}
  & \pmdg{0.0980}{0.001}
  & \pmdg{0.0981}{0.001}
  & \pmdg{0.105}{0.002}
  & \pmdg{0.096}{0.001}
  & \pmdg{0.099}{0.004}
\\ \hline
9\rule[-2.5mm]{0mm}{8mm}
  & \pmdg{0.0313}{0.0001}
  & \pmdg{0.0310}{0.001}
  & \pmdg{0.0313}{0.001}
  & \pmdg{0.0396}{0.001}
  & \pmdg{0.034}{0.001}
  & \pmdg{0.0386}{0.001}
\\ \hline
\end{tabular}

\vspace{0.5cm}
Table 4: {\it
Comparison for $g(\xi) = (1-\xi)^5$
}
\end{center}

\clearpage

\begin{center}
\begin{tabular}[h]{|c|c|c|c|c|c|c|}
\cline{2-7}
   \multicolumn{1}{c|}{\rule[-2.5mm]{0mm}{8mm}}
 & \multicolumn{3}{|c|}{leading order}
 & \multicolumn{3}{|c|}{next-to-leading order}
\\ \hline
  bin\rule[-2.5mm]{0mm}{8mm}
  & \makebox[2.2cm]{\tt DISASTER++}  
  & \makebox[2.2cm]{\tt MEPJET}  
  & \makebox[2.2cm]{\tt DISENT}  
  & \makebox[2.2cm]{\tt DISASTER++}  
  & \makebox[2.2cm]{\tt MEPJET}  
  & \makebox[2.2cm]{\tt DISENT}  
\\ \hline\hline
1\rule[-2.5mm]{0mm}{8mm}
  & \pmdg{46.1}{0.11}
  & \pmdg{46.2}{0.07}
  & \pmdg{46.2}{0.14}
  & \pmdg{49.4}{0.67}
  & \pmdg{58.8}{0.65}
  & \pmdg{47.8}{1.2}
\\ \hline
2\rule[-2.5mm]{0mm}{8mm}
  & \pmdg{23.8}{0.05}
  & \pmdg{23.8}{0.09}
  & \pmdg{23.8}{0.07}
  & \pmdg{30.6}{0.33}
  & \pmdg{31.4}{0.71}
  & \pmdg{29.0}{0.54}
\\ \hline
3\rule[-2.5mm]{0mm}{8mm}
  & \pmdg{7.28}{0.02}
  & \pmdg{7.28}{0.02}
  & \pmdg{7.29}{0.02}
  & \pmdg{11.2}{0.21}
  & \pmdg{11.0}{0.24}
  & \pmdg{11.4}{0.14}
\\ \hline
4\rule[-2.5mm]{0mm}{8mm}
  & \pmdg{42.4}{0.09}
  & \pmdg{42.3}{0.06}
  & \pmdg{42.3}{0.12}
  & \pmdg{38.4}{0.30}
  & \pmdg{41.9}{0.26}
  & \pmdg{38.4}{0.31}
\\ \hline
5\rule[-2.5mm]{0mm}{8mm}
  & \pmdg{23.9}{0.04}
  & \pmdg{23.9}{0.03}
  & \pmdg{23.8}{0.06}
  & \pmdg{24.8}{0.46}
  & \pmdg{24.2}{0.19}
  & \pmdg{23.9}{0.16}
\\ \hline
6\rule[-2.5mm]{0mm}{8mm}
  & \pmdg{7.31}{0.01}
  & \pmdg{7.30}{0.01}
  & \pmdg{7.27}{0.02}
  & \pmdg{8.11}{0.19}
  & \pmdg{8.04}{0.41}
  & \pmdg{8.24}{0.05}
\\ \hline
7\rule[-2.5mm]{0mm}{8mm}
  & \pmdg{20.3}{0.05}
  & \pmdg{20.3}{0.08}
  & \pmdg{20.3}{0.08}
  & \pmdg{23.3}{0.64}
  & \pmdg{25.1}{0.18}
  & \pmdg{22.4}{0.24}
\\ \hline
8\rule[-2.5mm]{0mm}{8mm}
  & \pmdg{15.4}{0.03}
  & \pmdg{15.4}{0.02}
  & \pmdg{15.4}{0.01}
  & \pmdg{18.6}{0.36}
  & \pmdg{18.3}{0.47}
  & \pmdg{18.4}{0.15}
\\ \hline
9\rule[-2.5mm]{0mm}{8mm}
  & \pmdg{4.87}{0.01}
  & \pmdg{4.86}{0.01}
  & \pmdg{4.87}{0.04}
  & \pmdg{6.47}{0.08}
  & \pmdg{6.38}{0.07}
  & \pmdg{6.41}{0.05}
\\ \hline
\end{tabular}

\vspace{0.5cm}
Table 5: {\it
Comparison for $q(\xi) = (1-\xi)^2$
}
\end{center}

\begin{center}
\begin{tabular}[h]{|c|c|c|c|c|c|c|}
\cline{2-7}
   \multicolumn{1}{c|}{\rule[-2.5mm]{0mm}{8mm}}
 & \multicolumn{3}{|c|}{leading order}
 & \multicolumn{3}{|c|}{next-to-leading order}
\\ \hline
  bin\rule[-2.5mm]{0mm}{8mm}
  & \makebox[2.2cm]{\tt DISASTER++}  
  & \makebox[2.2cm]{\tt MEPJET}  
  & \makebox[2.2cm]{\tt DISENT}  
  & \makebox[2.2cm]{\tt DISASTER++}  
  & \makebox[2.2cm]{\tt MEPJET}  
  & \makebox[2.2cm]{\tt DISENT}  
\\ \hline\hline
1\rule[-2.5mm]{0mm}{8mm}
  & \pmdg{6.24}{0.02}
  & \pmdg{6.22}{0.01}
  & \pmdg{6.21}{0.02}
  & \pmdg{6.73}{0.13}
  & \pmdg{8.94}{0.12}
  & \pmdg{6.67}{0.24}
\\ \hline
2\rule[-2.5mm]{0mm}{8mm}
  & \pmdg{3.59}{0.01}
  & \pmdg{3.58}{0.01}
  & \pmdg{3.57}{0.01}
  & \pmdg{4.77}{0.06}
  & \pmdg{5.24}{0.09}
  & \pmdg{4.43}{0.11}
\\ \hline
3\rule[-2.5mm]{0mm}{8mm}
  & \pmdg{1.18}{0.004}
  & \pmdg{1.18}{0.004}
  & \pmdg{1.18}{0.003}
  & \pmdg{1.93}{0.04}
  & \pmdg{1.89}{0.04}
  & \pmdg{1.86}{0.03}
\\ \hline
4\rule[-2.5mm]{0mm}{8mm}
  & \pmdg{2.65}{0.007}
  & \pmdg{2.65}{0.003}
  & \pmdg{2.65}{0.006}
  & \pmdg{1.13}{0.03}
  & \pmdg{1.66}{0.02}
  & \pmdg{0.94}{0.07}
\\ \hline
5\rule[-2.5mm]{0mm}{8mm}
  & \pmdg{1.62}{0.004}
  & \pmdg{1.61}{0.002}
  & \pmdg{1.61}{0.003}
  & \pmdg{1.04}{0.04}
  & \pmdg{1.09}{0.02}
  & \pmdg{0.993}{0.03}
\\ \hline
6\rule[-2.5mm]{0mm}{8mm}
  & \pmdg{0.535}{0.001}
  & \pmdg{0.534}{0.001}
  & \pmdg{0.533}{0.001}
  & \pmdg{0.433}{0.018}
  & \pmdg{0.412}{0.009}
  & \pmdg{0.430}{0.010}
\\ \hline
7\rule[-2.5mm]{0mm}{8mm}
  & \pmdg{0.452}{0.002}
  & \pmdg{0.452}{0.001}
  & \pmdg{0.451}{0.001}
  & \pmdg{0.221}{0.026}
  & \pmdg{0.292}{0.010}
  & \pmdg{0.129}{0.02}
\\ \hline
8\rule[-2.5mm]{0mm}{8mm}
  & \pmdg{0.398}{0.001}
  & \pmdg{0.398}{0.001}
  & \pmdg{0.397}{0.001}
  & \pmdg{0.298}{0.01}
  & \pmdg{0.271}{0.005}
  & \pmdg{0.237}{0.01}
\\ \hline
9\rule[-2.5mm]{0mm}{8mm}
  & \pmdg{0.136}{0.001}
  & \pmdg{0.135}{0.001}
  & \pmdg{0.135}{0.001}
  & \pmdg{0.130}{0.003}
  & \pmdg{0.109}{0.002}
  & \pmdg{0.120}{0.004}
\\ \hline
\end{tabular}

\vspace{0.5cm}
Table 6: {\it
Comparison for $g(\xi) = (1-\xi)^2$
}
\end{center}

\clearpage

\begin{center}
\begin{tabular}[h]{|c|c|c|c|c|c|c|}
\cline{2-7}
   \multicolumn{1}{c|}{\rule[-2.5mm]{0mm}{8mm}}
 & \multicolumn{3}{|c|}{leading order}
 & \multicolumn{3}{|c|}{next-to-leading order}
\\ \hline
  bin\rule[-2.5mm]{0mm}{8mm}
  & \makebox[2.2cm]{\tt DISASTER++}  
  & \makebox[2.2cm]{\tt MEPJET}  
  & \makebox[2.2cm]{\tt DISENT}  
  & \makebox[2.2cm]{\tt DISASTER++}  
  & \makebox[2.2cm]{\tt MEPJET}  
  & \makebox[2.2cm]{\tt DISENT}  
\\ \hline\hline
1\rule[-2.5mm]{0mm}{8mm}
  & \pmdg{50.6}{0.12}
  & \pmdg{50.7}{0.13}
  & \pmdg{50.7}{0.15}
  & \pmdg{58.6}{1.29}
  & \pmdg{72.9}{1.56}
  & \pmdg{54.7}{2.1}
\\ \hline
2\rule[-2.5mm]{0mm}{8mm}
  & \pmdg{27.1}{0.05}
  & \pmdg{27.1}{0.16}
  & \pmdg{27.0}{0.07}
  & \pmdg{36.4}{0.57}
  & \pmdg{40.0}{0.84}
  & \pmdg{34.9}{1.0}
\\ \hline
3\rule[-2.5mm]{0mm}{8mm}
  & \pmdg{8.51}{0.02}
  & \pmdg{8.51}{0.02}
  & \pmdg{8.52}{0.02}
  & \pmdg{13.8}{0.35}
  & \pmdg{13.3}{0.43}
  & \pmdg{13.9}{0.2}
\\ \hline
4\rule[-2.5mm]{0mm}{8mm}
  & \pmdg{49.8}{0.10}
  & \pmdg{49.7}{0.05}
  & \pmdg{49.6}{0.14}
  & \pmdg{41.2}{0.55}
  & \pmdg{47.2}{0.91}
  & \pmdg{41.9}{0.38}
\\ \hline
5\rule[-2.5mm]{0mm}{8mm}
  & \pmdg{29.0}{0.05}
  & \pmdg{29.0}{0.03}
  & \pmdg{28.8}{0.07}
  & \pmdg{27.3}{0.52}
  & \pmdg{28.2}{0.42}
  & \pmdg{26.4}{0.19}
\\ \hline
6\rule[-2.5mm]{0mm}{8mm}
  & \pmdg{9.09}{0.01}
  & \pmdg{9.07}{0.01}
  & \pmdg{9.04}{0.02}
  & \pmdg{9.58}{0.06}
  & \pmdg{9.16}{0.15}
  & \pmdg{9.54}{0.06}
\\ \hline
7\rule[-2.5mm]{0mm}{8mm}
  & \pmdg{30.6}{0.08}
  & \pmdg{30.5}{0.04}
  & \pmdg{30.5}{0.12}
  & \pmdg{32.0}{0.34}
  & \pmdg{36.3}{0.59}
  & \pmdg{32.4}{0.52}
\\ \hline
8\rule[-2.5mm]{0mm}{8mm}
  & \pmdg{24.3}{0.04}
  & \pmdg{24.3}{0.03}
  & \pmdg{24.3}{0.07}
  & \pmdg{27.6}{0.56}
  & \pmdg{28.4}{0.35}
  & \pmdg{27.6}{0.21}
\\ \hline
9\rule[-2.5mm]{0mm}{8mm}
  & \pmdg{7.88}{0.01}
  & \pmdg{7.86}{0.01}
  & \pmdg{7.87}{0.02}
  & \pmdg{9.63}{0.21}
  & \pmdg{9.50}{0.15}
  & \pmdg{9.47}{0.06}
\\ \hline
\end{tabular}

\vspace{0.5cm}
Table 7: {\it
Comparison for $q(\xi) = (1-\xi)$
}
\end{center}

\begin{center}
\begin{tabular}[h]{|c|c|c|c|c|c|c|}
\cline{2-7}
   \multicolumn{1}{c|}{\rule[-2.5mm]{0mm}{8mm}}
 & \multicolumn{3}{|c|}{leading order}
 & \multicolumn{3}{|c|}{next-to-leading order}
\\ \hline
  bin\rule[-2.5mm]{0mm}{8mm}
  & \makebox[2.2cm]{\tt DISASTER++}  
  & \makebox[2.2cm]{\tt MEPJET}  
  & \makebox[2.2cm]{\tt DISENT}  
  & \makebox[2.2cm]{\tt DISASTER++}  
  & \makebox[2.2cm]{\tt MEPJET}  
  & \makebox[2.2cm]{\tt DISENT}  
\\ \hline\hline
1\rule[-2.5mm]{0mm}{8mm}
  & \pmdg{6.84}{0.02}
  & \pmdg{6.84}{0.01}
  & \pmdg{6.82}{0.02}
  & \pmdg{8.20}{0.25}
  & \pmdg{11.6}{0.14}
  & \pmdg{8.26}{0.45}
\\ \hline
2\rule[-2.5mm]{0mm}{8mm}
  & \pmdg{4.09}{0.01}
  & \pmdg{4.07}{0.01}
  & \pmdg{4.07}{0.01}
  & \pmdg{5.70}{0.11}
  & \pmdg{6.69}{0.16}
  & \pmdg{5.68}{0.17}
\\ \hline
3\rule[-2.5mm]{0mm}{8mm}
  & \pmdg{1.39}{0.004}
  & \pmdg{1.39}{0.005}
  & \pmdg{1.39}{0.003}
  & \pmdg{2.41}{0.07}
  & \pmdg{2.33}{0.05}
  & \pmdg{2.34}{0.05}
\\ \hline
4\rule[-2.5mm]{0mm}{8mm}
  & \pmdg{3.19}{0.01}
  & \pmdg{3.19}{0.01}
  & \pmdg{3.19}{0.01}
  & \pmdg{0.686}{0.09}
  & \pmdg{1.65}{0.03}
  & \pmdg{0.691}{0.10}
\\ \hline
5\rule[-2.5mm]{0mm}{8mm}
  & \pmdg{2.06}{0.005}
  & \pmdg{2.06}{0.002}
  & \pmdg{2.05}{0.003}
  & \pmdg{1.00}{0.08}
  & \pmdg{1.14}{0.03}
  & \pmdg{0.866}{0.05}
\\ \hline
6\rule[-2.5mm]{0mm}{8mm}
  & \pmdg{0.711}{0.001}
  & \pmdg{0.710}{0.001}
  & \pmdg{0.709}{0.001}
  & \pmdg{0.500}{0.006}
  & \pmdg{0.471}{0.01}
  & \pmdg{0.442}{0.017}
\\ \hline
7\rule[-2.5mm]{0mm}{8mm}
  & \pmdg{0.712}{0.003}
  & \pmdg{0.711}{0.001}
  & \pmdg{0.710}{0.002}
  & \pmdg{0.157}{0.026}
  & \pmdg{0.373}{0.008}
  & \pmdg{0.082}{0.038}
\\ \hline
8\rule[-2.5mm]{0mm}{8mm}
  & \pmdg{0.692}{0.002}
  & \pmdg{0.690}{0.001}
  & \pmdg{0.690}{0.001}
  & \pmdg{0.411}{0.020}
  & \pmdg{0.408}{0.022}
  & \pmdg{0.340}{0.023}
\\ \hline
9\rule[-2.5mm]{0mm}{8mm}
  & \pmdg{0.245}{0.001}
  & \pmdg{0.245}{0.001}
  & \pmdg{0.245}{0.001}
  & \pmdg{0.194}{0.012}
  & \pmdg{0.172}{0.007}
  & \pmdg{0.161}{0.008}
\\ \hline
\end{tabular}

\vspace{0.5cm}
Table 8: {\it
Comparison for $g(\xi) = (1-\xi)$
}
\end{center}

\end{appendix}

\clearpage

\newcommand{\bibitema}[1]{\bibitem{#1}}

\end{document}